# Steganography: A Secure way for Transmission in Wireless Sensor Networks

**Khan Muhammad**

Department of Computer Science, Islamia College Peshawar, Pakistan

[E-mail: khan.muhammad.icp@gmail.com]

## Abstract

Addressing the security concerns in wireless sensor networks (WSN) is a challenging task, which has attracted the attention of many researchers from the last few decades. Researchers have presented various schemes in WSN, addressing the problems of processing, bandwidth, load balancing, and efficient routing. However, little work has been done on security aspects of WSN. In a typical WSN network, the tiny nodes installed on different locations sense the surrounding environment, send the collected data to their neighbors, which in turn is forwarded to a sink node. The sink node aggregate the data received from different sensors and send it to the base station for further processing and necessary actions. In highly critical sensor networks such as military and law enforcement agencies networks, the transmission of such aggregated data via the public network "Internet" is very sensitive and vulnerable to various attacks and risks. Therefore, this paper provides a solution for addressing these security issues based on steganography, where the aggregated data can be embedded as a secret message inside an innocent-looking cover image. The stego image containing the embedded data can be then sent to fusion center using Internet. At the fusion center, the hidden data is extracted from the image, the required processing is performed and decision is taken accordingly. Experimentally, the proposed method is evaluated by objective analysis using peak signal-to noise ratio (PSNR), mean square error (MSE), normalized cross correlation (NCC), and structural similarity index metric (SSIM), providing promising results in terms of security and image quality, thus validating its superiority.

# 1. Introduction

A WSN contains multiple distributed sensors for monitoring different environmental conditions like pressure, temperature, motion, sound, and images[1]. The data sensed by these sensors is transmitted to the fusion center for further necessary action in WSNs. This transmission of data to the base station is vulnerable to many risks. For example, an attacker can easily modify the actual data captured by sensor nodes during its transmission because WSNs are almost open networks with no strong security considerations like wired networks. When the fusion center receive wrong data, the decision taken on the basis of this data will be definitely wrong. This wrong decision can lead a law enforcement authority towards horrible destruction. Because of these reasons, maintaining the security in WSN is very important. Although more research is done on the critical constraints of WSN such as power consumption, bandwidth and memory limitations, and computational powers, yet security in WSN is also an open challenge which needs to be addressed especially in sensitive sensor networks of military and law enforcement agencies[2-7].

The most important and challenging issues that require urgent solutions in different advanced multimedia applications of wireless sensor networks include localization of sensors based on images, object tracking using sensor image processing, aggregation of images in sensor nodes, image processing for minimizing the computations, bandwidth and energy limitations[8, 9], coverage of object view-angle using visual sensor networks, image processing for network security in WSN[10, 11], pre-processing inside WSN like image compression, and efficient and effective capturing of video and images [6, 12, 13].

## 1.1 Structure of Wireless Sensor Node

A wireless sensor node consists of four main components and some optional components. The optional components depend on the type of application. These components are described below and depicted in Fig. 1.

i) Each sensor node occupies a specific sensing unit that contains one or multiple sensors plus an A/D converter used for data acquirement.

ii) Each sensor node has a central processing unit plus some specific amount of memory for storage of intermediate results and other data and a micro-controller.

iii) An RF unit used for communication of data using wireless media.

iv) A special unit for providing power to sensor nodes.

v) A system for position and location determination. (optional component)

vi) A unit known as mobilizer for configuration and location changing. (optional component)

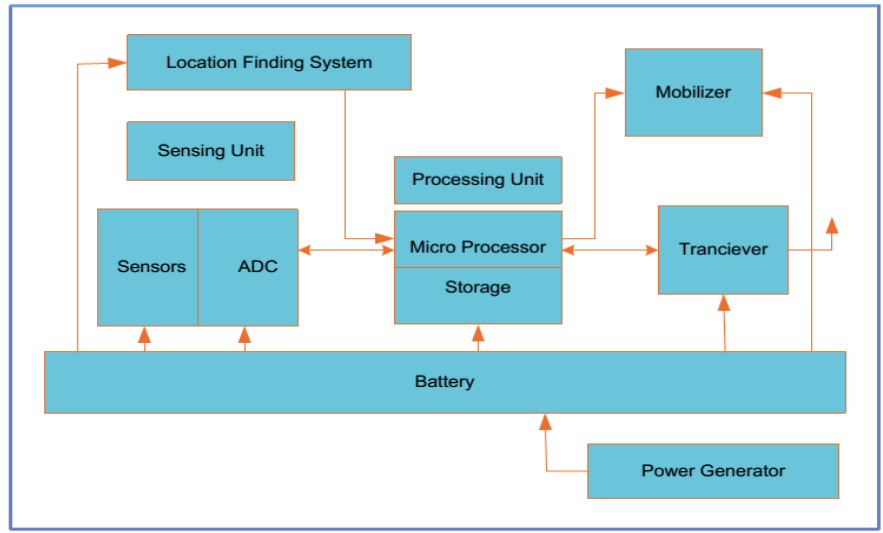

Fig. 1: Structure of a Wireless Sensor Node

WSNs can be used for innumerable applications in different areas like smart farming to monitor the environment for effective utilization of water and land resources, target tracking, under-water sensing, traffic monitoring and enforcement, medical diagnosis[14, 15], multi-scale tracking[16], image change detection, smart parking, networked gamming, habitat monitoring, vineyard monitoring, remote sensing[17], and smart video and audio surveillance[2, 18-23].

**1.2 Steganography**

The word steganography is a Greek origin word meaning "Protected Writing". It can be defined as, the process during which secret information is embedded inside a carrier object such that it cannot be detected by the human visual system (HVS)[24]. Requirements of steganography include a carrier object, secret data, embedding algorithm, and sometimes a secret key and encryption algorithm to increase the security[25]. Its application involves the exchange of top secret information between international governments and defense organizations, medical imaging, online banking security, smart identity card security, online voting security, and secure exchange of data sensed by wireless sensor nodes in WSNs[26, 27]. In negative sense it can be used for sending viruses and Trojan horses and provides a best method to be used by terrorists and criminals for their confidential communication[28, 29]. The different techniques used for steganography on the basis of carrier object used at the time of embedding are described as follows [30-32]:

- Text Steganography
- Image Steganography
- Video Steganography
- Audio Steganography
- Network Steganography

**1.3 Classification of Steganographic Techniques**

The steganographic techniques can be broadly classified into two main categories depending upon the way these techniques modify the image pixels to hide secret data such as secret messages, secret images of map, and different parameters sensed by wireless sensor nodes. These categories are depicted in Fig. 2 and are briefly described as follows.

**A. Spatial Domain Techniques**

Spatial domain techniques directly modify the carrier image pixels in order to hide the secret information. These techniques possess high payload capacity but are vulnerable to statistical attacks like chi-square test, image rotation, compression[33]. Examples of spatial domain techniques include least significant bit (LSB), edges based embedding (EBE), pixel value differencing (PVD), pixel indicator technique (PIT) and gray-level modification (GLM)[30].

**B. Transform Domain Techniques**

Transform domain techniques convert the carrier image from spatial domain into transform domain with the help of different transforms like discrete cosine transform (DCT), discrete wavelet transform (DWT), and discrete Fourier transform (DFD) and alter the coefficients of resultant image to embed the secret information. At the end, the image is again shifted to spatial domain by taking inverse DCT, DWT or DFD of the resultant image. These techniques possess lower payload but are more robust and have better resistance against different statistical attacks. Examples of transform domain techniques are integer contour transform techniques (ICTT), DWT techniques, DCT techniques, Arnold transform technique (ATT) and DFD techniques[30].

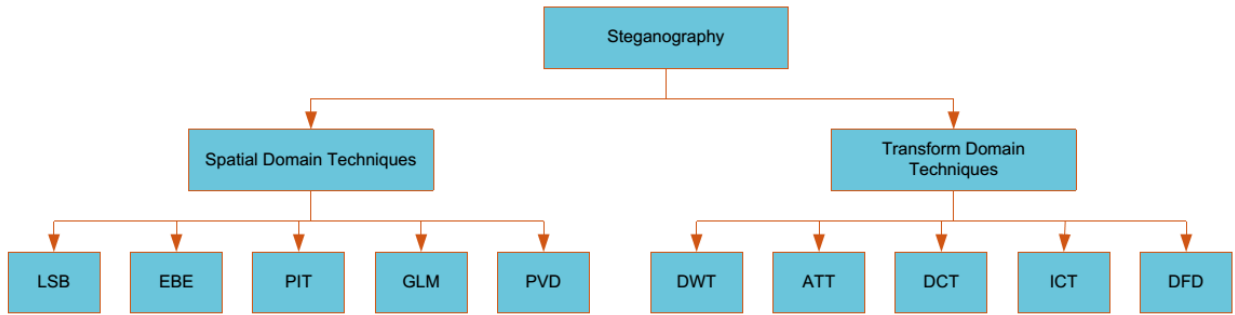

Fig. 2: Classifications of steganographic techniques

In this paper, a steganography based scheme is proposed to address the security issues in WSN. The major contributions of this research work are summarized as follows:

   i.   We identify a real-world problem in WSN in terms of security and propose a solution using image steganography, where the aggregated sensed data is embedded in images, keeping them visually and statistically invisible. To the best of our knowledge, our solution is the first inclination towards security in WSN using steganography.

ii. The aggregated data is embedded in cover images using cyclic LSB substitution method, producing relatively better stego images with acceptable visual quality.

iii. The proposed scheme is computationally inexpensive, balancing the visual quality and computational complexity, making it more suitable for real-time processing in WSN.

The rest of the paper is organized as follows. In section 2 several related approaches are discussed, whose limitations led us towards the current proposed work. Section 3 describes the proposed work. Experimental results and discussion is detailed in section 4 and section 5 concludes the paper.

## 2. Literature Review

The rapid development and advancement in wireless sensor networks allow us to use relatively inexpensive nodes having camera sensors and are connected with one another and one or multiple central servers via a network. These sensor nodes vary in size and cost depending upon its complexity. Size and cost constraints of sensor nodes result in corresponding constraints on resources such as energy, memory, computational speed, and communication bandwidth. To efficiently use these resources different approaches have been used.

Pathan, Lee, and Hong present a detailed discussion on the major challenges and well known attacks on WSN in [34]. The open challenges of WSN are correct data collection, trust management, secure data aggregation, communication, and computation load of resource restricted devices. Different types of attacks on WSN such as wormhole attack, denial of service (DOS) attack, hello flood attack, sybil attack, selecting forwarding, and sinkhole attack are critically discussed by the authors that demand for urgent solutions from the researchers of WSN.

Muhammad Atique et al. [35] proposed a secure routing technique to detect false reports and gray-hole attack by making use of statistical en-route filtering (SEF) in order to increase the security level in WSN. To further improve the security and reduce the energy consumption during transmission of sensed data, the authors implemented elliptical curve cryptography (ECC). The proposed method provided promising results in terms of security and created several barriers in the way of an attacker.

A detailed critical discussion is presented in [36] on the different aspects of WSN security issues including its design, context, integrity, authenticity, confidentiality, and algorithms to cope with these security issues. The authors proposed a practical algorithm for data security (PADS) that calculates a message authentication code (MAC) of 4 byte with the help of packet's static part. This MAC is appended with the data and a time synced key is generated via a secret key that is shared between the communicating bodies. By this way, it is very difficult for an attacker to break down the encryption because he needs to be time synced with the network which may be difficult for a malicious user. At the end, the authors briefly describe self-originating WSN (SOWSN) which also provides distinctive security features in WSN.

Wibhada et al. [37] proposed a new node replication detection method known as area-based clustering detection (ABCD) method in order to handle the problem of node replication attack in WSN. The proposed method reduces the communication overhead and provide high correct detection rate when compared to line-selected multicast (LSM) technique. Similarly this new ABCD technique also reduces the number of stored messages and prolongs the network lifetime as compared to centralized method.

## 3. The Proposed Method

In this work, a new approach is proposed to handle the security problems in WSN during data transmission by using steganography. In critical sensor networks such as military systems, video surveillance and multi-scale tracking systems, the sensor nodes continuously sense the surrounding environment and send the collected data to a sink node via multi-hop communication. So each sensor node plays two different roles: data collection and acting as a rely point. The sink node is then responsible to send the received data to base station for further processing and necessary action. The proposed method copes with these security issues of data transmission from one sensor node to another and finally to base station in WSNs. The method used for data hiding is cyclic LSB substitution method. This method hides the sensitive data collected by sensor nodes in a carrier color image such that it cannot be detected by the HVS. At the base station, the embedded sensitive data is extracted from the stego image and further processing is performed on it accordingly. Although this approach may require more bandwidth and processing but in most sensitive WSNs such as atomic energy networks and intelligent agencies networks, these issues are acceptable as they cannot compromise on security. Due to this reason, the proposed model of steganography plays a vital role in coping with security issues of WSNs. The next two subsections present the embedding and extraction algorithms used for hiding of sensitive data collected by sensor nodes.

### 3.1 Embedding Algorithm

*Input:* Color Image and sensitive data sensed by sensor nodes

*Output:* Stego Image

Step 1: Take the cover color image and sensitive data.

Step 2: Separate the RED, GREEN and BLUE planes from the cover image.

Step 3: Convert sensitive data into 1-D array of bits.

Step 4: Set indicator = 1 initially.

Step 5: If indicator = 1

    Replace the LSB of RED channel with sensitive bit

  Else if indicator = 2

    Replace the LSB of GREEN channel with sensitive bit

  Else if indicator = 3

         Replace the LSB of BLUE channel with sensitive bit

  End

Step 6: If indicator = 3

      Set indicator = 1;

  End

Step 7: Repeat Step 5 and Step 6 until all sensitive data bits are embedded.

Step 8: Combine all three planes to form the resultant stego image.

### 3.2 Extraction Algorithm

*Input:* Stego Image

*Output:* Secret data

Step 1: Take the stego image and separate the RED, GREEN and BLUE planes from it.

Step 2: Set indicator = 1 initially.

Step 3: If indicator = 1

    Extract the LSB of RED channel.

  Else if indicator = 2

    Extract the LSB of GREEN channel.

  Else if indicator = 3

    Extract the LSB of BLUE channel.

  End

Step 4: If indicator = 3

    Set indicator = 1;

  End

Step 5: Repeat Step 3 and Step 4 until all secret data bits are extracted.

Step 6: Convert the extracted secret bits into its original secret data format.

## 4. Experimental Results and Discussion

The proposed method is simulated using MATLAB R2013a. For experiments, different amount of sensitive data is embedded in different standard color images of same and variable dimensions. The standard color images used for experimental purposes include lena.png, baboon.png, masjid.png, and trees.tiff. The proposed method is evaluated experimentally from three different viewpoints. First of all, same amount of sensitive data (8KB) is embedded in different standard images of same dimensions (256×256). Secondly, variable amount of secret data is hidden in the same image (baboon) of same dimension (256×256). The last viewpoint is to store same amount of cipher in same image (baboon) of different dimensions.

Objective analysis using PSNR, MSE, NCC, and SSIM is performed on the proposed method to evaluate its performance[38]. PSNR is a statistical image quality assessment standard used for measuring the obvious distortion between stego and cover image[39]. The PSNR is measured in decibels (dB)[40]. PSNR values below 30dB show low quality of stego images and hence it brings noticeable changes in stego images which can be seen by naked eyes[41, 42]. To achieve good quality of stego images, PSNR value must be 40dB or above than 40bB[43]. MSE is used to calculate the error between the original and stego image[44]. NCC shows how strongly the stego image is correlated with the original image[45]. The value of NCC is between 1 and 0. SSIM is another IQAM used for measuring the noticeable distortion between the host and stego image[14]. The PSNR, MSE, NCC, and SSIM are calculated by the following formulae.

$$\text{PSNR} = 10\log_{10}\left(\frac{C_{\max\ 2}}{\text{MSE}}\right) \quad (1)$$

$$\text{MSE} = \frac{1}{MN}\sum_{x=1}^{M}\sum_{y=1}^{N}(S_{xy} - C_{xy}) \quad (2)$$

$$\text{NCC} = \frac{\sum_{x=1}^{M}\sum_{y=1}^{N}(S(x,y)*C(x,y))}{\sum_{x=1}^{M}\sum_{y=1}^{N}S(x,y)^2} \qquad (3)$$

$$\text{SSIM}(X,Y) = \frac{(2\mu_x\mu_y+C_1)(2\sigma_{xy}+C_2)}{(\mu_x^2+\mu_y^2+C_1)(\sigma_x^2+\sigma_y^2+C_2)} \qquad (4)$$

Here $C_{max}$ is the maximum pixel intensity in both images, M and N are image dimensions, x and y are loop variables, S represents stego image and C shows the cover image. The incurred results of the proposed method are tabulated in Table 1-3.

Table 1: Experimental results of proposed method using perspective1

| Serial# | Image Name | PSNR (dB) | MSE | NCC | SSIM |
|---|---|---|---|---|---|
| 1 | couple | 55.9144 | 0 | 0.9999 | 0.9985 |
| 2 | house | 51.1776 | 0 | 0.9999 | 0.999 |
| 3 | Lena | 55.9211 | 0 | 1 | 0.9989 |
| 4 | f16jet | 47.4852 | 0.0001 | 0.9997 | 0.9985 |
| 5 | baboon | 48.9531 | 0.0001 | 0.9998 | 0.9993 |
| 6 | house | 51.1776 | 0 | 0.9999 | 0.999 |
| 7 | Moon | 47.4921 | 0.0001 | 0.9998 | 0.9986 |
| 8 | army | 55.9201 | 0 | 1 | 0.9991 |
| 9 | Scene3 | 55.9306 | 0 | 0.9999 | 0.9994 |
| 10 | rose | 55.9337 | 0 | 1 | 0.9991 |
| **Average of 20 images** | | **49.8018** | **0.0001** | **0.99987** | **0.998555** |

Table 1 shows the experimental results of the proposed method using perspective1 based on PSNR, MSE, NCC, and SSIM. In this experiment, a message of size 8KB is embedded in 20 standard color images of dimension 256 by 256 pixels and the results are tabulated in Table 1 using various IQAMs. Ten (10) images are shown with their names and the last row shows the average value for each metric over 20 images. The high values of PSNR, SSIM and NCC of Table 1 show the better imperceptibility of the proposed method.

Table 2: Experimental results of proposed method using perspective2

| Image Name | Secret data (KBs) | Cipher size in bytes | PSNR | MSE | NCC | SSIM |
|---|---|---|---|---|---|---|
| **Lena with resolution 256×256** | 2 | 2406 | 61.7772 | 0.0432 | 1 | 0.9997 |
| | 4 | 4177 | 58.7552 | 0.0866 | 1 | 0.9995 |
| | 6 | 6499 | 56.9562 | 0.1311 | 1 | 0.9992 |
| | 8 | 8192 | 55.9211 | 0.1663 | 1 | 0.9989 |
| | Average | | **58.3524** | **0.1068** | **1** | **0.999325** |
| **Baboon image with dimension 256×256** | 2 | 2406 | 49.6451 | 0.7056 | 0.9999 | 0.9996 |
| | 4 | 4177 | 49.3853 | 0.7491 | 0.9999 | 0.9994 |
| | 6 | 6499 | 49.1359 | 0.7934 | 0.9998 | 0.9993 |
| | 8 | 8192 | 48.9531 | 0.8275 | 0.9998 | 0.9993 |
| | Average | | **49.2799** | **0.7689** | **0.99985** | **0.9994** |
| **Building image with dimension 256×256** | 2 | 2406 | 61.6587 | 0.0444 | 1 | 0.9995 |
| | 4 | 4177 | 58.631 | 0.0891 | 1 | 0.9991 |
| | 6 | 6499 | 56.8991 | 0.1328 | 1 | 0.9987 |
| | 8 | 8192 | 55.8999 | 0.1671 | 1 | 0.9984 |
| | Average | | **58.2722** | **0.10835** | **1** | **0.998925** |
| **House image with dimension 256×256** | 2 | 2406 | 52.3894 | 0.3751 | 0.9999 | 0.9998 |
| | 4 | 4177 | 51.9059 | 0.4193 | 0.9999 | 0.9995 |
| | 6 | 6499 | 51.4802 | 0.4624 | 0.9999 | 0.9992 |
| | 8 | 8192 | 51.1776 | 0.4958 | 0.9999 | 0.999 |
| | Average | | **51.7383** | **0.43815** | **0.9999** | **0.999375** |

Table 2 shows the experimental results of the proposed method based on PSNR, NCC, MSE, and SSIM using perspective2. In this experiment, different amount of data is embedded in a few standard sample cover images while keeping the resolution of the images constant i.e. 256 by 256 pixels. The average values based on each metric are shown in bold form for every sample image. By noticing the average values, one can see that there is a little decrease in the values of the given metrics as compared to increase in secret data.

Table 3: Experimental results of proposed method using perspective3

| Image Name | Image dimensions | PSNR | MSE | NCC | SSIM |
|---|---|---|---|---|---|
| Lena image | 128×128 | 42.1215 | 3.9896 | 0.999 | 0.9989 |
| | 256×256 | 47.4888 | 1.1593 | 0.9997 | 0.9983 |
| | 512×512 | 48.7424 | 0.8686 | 0.9998 | 0.9997 |
| | 1024×1024 | 49.858 | 0.6719 | 0.9998 | 0.9999 |
| | **Average** | **47.05268** | **1.67235** | **0.999575** | **0.9992** |
| Building image | 128×128 | 64.7241 | 0.0219 | 1 | 0.9998 |
| | 256×256 | 47.4889 | 1.1593 | 0.9997 | 0.9979 |
| | 512×512 | 47.9868 | 1.0337 | 0.9997 | 0.9996 |
| | 1024×1024 | 48.9026 | 0.8372 | 0.9997 | 0.9998 |
| | **Average** | **52.2756** | **0.763025** | **0.999775** | **0.999275** |
| Baboon image | 128×128 | 64.9432 | 0.0208 | 1 | 1 |
| | 256×256 | 55.916 | 0.1665 | 0.9999 | 0.9992 |
| | 512×512 | 61.9202 | 0.0418 | 1 | 1 |
| | 1024×1024 | 67.9482 | 0.0104 | 1 | 1 |
| | **Average** | **62.6819** | **0.059875** | **0.999975** | **0.9998** |
| House image | 128×128 | 64.8926 | 0.0211 | 1 | 1 |
| | 256×256 | 41.0315 | 5.1278 | 0.9993 | 0.9983 |
| | 512×512 | 42.1893 | 3.9279 | 0.9994 | 0.9997 |
| | 1024×1024 | 43.1444 | 3.1524 | 0.9994 | 0.9998 |
| | **Average** | **47.81445** | **3.0573** | **0.999525** | **0.99945** |

Table 3 shows the experimental results of the proposed approach using perspective3 based on different IQAMs. In this experiment, the size of secret data is kept constant while the dimension of images is variable. The quantitative results based on PSNR, NCC, MSE, and SSIM along with the bold face average values show that the proposed scheme provide better results and do not bring obvious deformation in the stego images after embedding data.

The proposed method is also evaluated using subjective analysis. In subjective analysis, the HVS is used to check that how much the stego image is similar to the original image. The more the stego image is similar to the original cover image; the more the visual quality of the stego image will be better and vice versa. The upshots of cover and stego images of the proposed algorithm are shown in Fig. 3. It is clear from Fig. 3 that the stego images are almost same as the cover images and there is no visible deformation in the stego images which shows the high imperceptibility of the proposed technique.

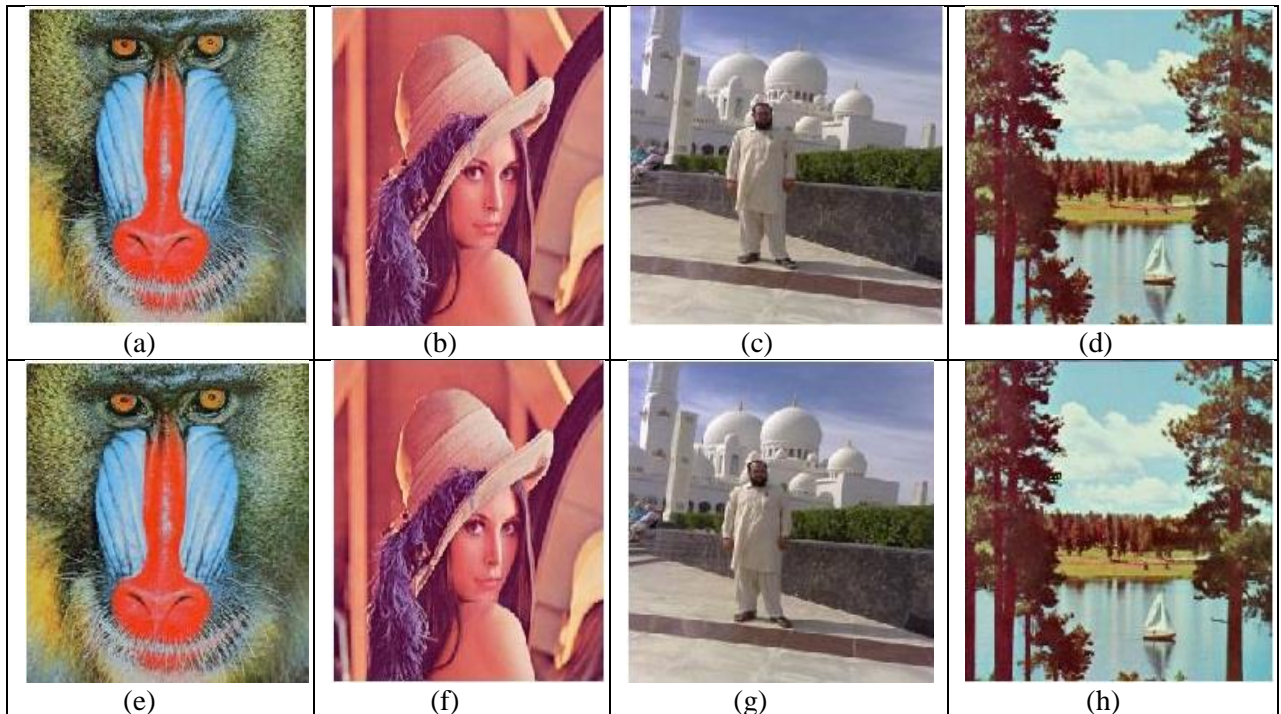

Fig. 3: Subjective evaluation using HVS. First row shows 256×256 sized cover images (a) Baboon (b) Lena (c) Masjid (d) Trees. Second row shows 256×256 sized stego images (e) Baboon (f) Lena (g) Masjid (h) Trees

## 5. Conclusion

In this paper, we proposed a new approach to solve the security issues of sensitive data transmission in WSNs by using steganography. Steganography embeds the data collected by wireless sensor node into an image in order to minimize different fraudulent behaviors. The performance of the proposed method is evaluated by PSNR, MSE, NCC, and SSIM. An encouraging finding of the proposed technique is the average values of PSNR (PSNR>=47dB), NCC (NCC>=0.999), and SSIM (SSIM>=0.999) found in all cases with same and variable amount of cipher with different standard images of same and different dimensions. The proposed approach provides secure transmission of sensed data and is a good solution to be adopted by military and law-enforcement agencies for critical security applications.